\def\be{\begin{eqnarray}}
\def\ee{\end{eqnarray}}

\documentstyle[12pt,psfig]{article}

\begin{document}
\thispagestyle{empty}

\begin{flushright}
BNL-HET-98/21\\
TTP 98--22\\
hep-ph/9806244
\end{flushright}
\vspace{0.6cm}
\boldmath
\begin{center}
\Large\bf Two-loop QCD corrections to top quark width
\end{center}
\unboldmath
\vspace{0.8cm}

\begin{center}

{\large Andrzej Czarnecki}\\
{\sl Physics Department, Brookhaven National Laboratory,\\
Upton, NY 11973}

\vspace*{5mm}

{\large Kirill Melnikov}\\
{\sl Institut f\"{u}r Theoretische Teilchenphysik,Universit\"at Karlsruhe,}\\
{\sl     D--76128 Karlsruhe, Germany} \\
\end{center}

\vfill

\begin{abstract}
We present ${\cal O}(\alpha_s^2)$ corrections to the decay $t
\to bW$ in the limit of a very large top quark mass, $m_t\gg m_W$.  We
find that the ${\cal O}(\alpha_s^2)$ effects decrease the top quark
decay width $\Gamma_t$ by about 2\%:
$\Gamma_t=\Gamma_0(1-0.8\alpha_s(m_t)-1.7\alpha_s^2)$.  The complete
corrections are smaller by about 24\% than their estimate based on the
BLM effects ${\cal O}(\beta_0\alpha_s^2)$.  We explain how to compute
a new type of diagrams which contribute to $\Gamma(t \to b W)$ at the
${\cal O}(\alpha _s^2)$ level. 
\end{abstract}

\vfill
\vfill
\vfill

\newpage
\section{Introduction}

Because of its very large mass, the top quark is an interesting object
for precise studies and searches for possible 
``new physics.''  In consequence
of the large $m_t$, top lifetime is very short. With good accuracy the
top decay rate $\Gamma_t$ is proportional to the third power of the
top mass and equals approximately $1.5$ GeV for $m_t=175$ GeV.  This
large width makes the top quark behave almost like a {\it free} quark,
a unique situation in QCD \cite{APP}.  
The reason for this is that the time scale
$\tau _W$ of the weak top quark decay 
$$
\tau _W \approx \left [  \frac {G_F m_t^3}{8\sqrt{2}\pi} \right ]^{-1},
$$
is much shorter than the time scale of the non-perturbative 
QCD effects, $1/\Lambda_{\rm QCD}$.

Within the Standard Model (SM) the reaction $t \to b W$ is the dominant top
quark decay channel.  However, in most extensions of the standard
model (SM) there are additional decay channels (e.g.~$t \to b H^+$ or
decays into final states including supersymmetric particles).  
Since the SM prediction for the top quark lifetime can be
given with high precision, its measurements can  in principle shed
light on the new physics contributions to top decay.  

%It has also been speculated that the proximity of m_t to the
%electroweak scale suggests a special role that the top quark plays in
%electroweak symmetry breaking.  

The fact that the life time of the top quark is so small
has rather interesting consequences for the top production 
in $e^+e^-$ annihilation.
Because the nonperturbative effects are suppressed by the large
$\Gamma_t$, threshold production cross section of $t \bar t$ in $e^+
e^-$ collisions can be predicted rather reliably.  Clean signature and
a rich physics potential \cite{topphys} make the investigation of this
process in the threshold region one of the priorities at the Next
Linear Collider (NLC).  Many studies have recently been performed in
order to describe this reaction with high precision (see
e.g.~\cite{Jezabek:1994qs,HT,MeYel}). 

The total cross section for $e^+e^- \to t \bar t$ in the threshold
region is determined by the equation:
\begin{equation}
\sigma _t (s) \sim {\rm Im }\; G(E+i \Gamma_t;0,0),
\end{equation}
where $s$ is the center of mass energy squared, $E \equiv \sqrt{s} -
2m_t$, and $G(E;0,0)$ is the value  at the origin of the
Schr\"odinger equation Green function describing the non--relativistic
$t \bar t$ system. Because of the $\sigma_t(s)$ dependence on
$\Gamma_t$, a precise prediction for the cross section requires a
matching accuracy in $\Gamma_t$.  This is true not only for the total
cross section but also for other top threshold observables, for
example top momentum distribution.

The analysis of $\sigma_t(s)$ has been recently extended to the
next-to-next-to-leading order accuracy \cite{HT,MeYel}.  That analysis
was made possible in part by the recently calculated two-loop
corrections to the QCD potential \cite{Peter:1997ig}, whose
influence on the threshold cross section was investigated in
\cite{Jezabek:1998pj}.  On the other hand, the ${\cal
O}(\alpha_s^2)$ corrections to the width were not known at that time
and were not included into those studies.  The purpose of the present
paper is to provide $\Gamma_t$ including ${\cal O}(\alpha_s^2)$
effects.

The direct measurement of the top quark decay width at NLC is
difficult. At the moment a determination of the top quark width based
on the measurement of the forward-backward asymmetry of $t$ quarks in
the threshold region in $e^+ e^-$ collisions is considered to be the
most promising option.  It is believed that the accuracy of $10-20\%$
can be obtained in such a measurement \cite{topphys}.  More optimistic
estimates of 5\% accuracy have also been published
\cite{Fujii}.  Still better accuracy might be obtained at a future
muon collider. 

Because of the importance of the top quark width, much effort has been
invested into studies of radiative corrections to its decays.  Here we
briefly summarize the results obtained in the SM.  For a
more extensive discussion and a summary of calculations in some
extensions of the SM we refer the reader to
ref.~\cite{Kuhn:1996ug,Jezabek:1993wk}, where also corrections to
various differential decay distributions are presented.

QCD corrections to heavy quark decays were first studied in an
effective Fermi-like theory, valid for quarks much lighter than the
$W$ \cite{Cabibbo78}.  For the semileptonic
decays such calculations were technically similar to the muon decay
case \cite{beh56}.  After it had become clear that the top
quark is heavy, QCD corrections were calculated in \cite{jk2} with
full $W$ propagator taken into account.  These results are also
applicable to the process $t\to bW$ which is now considered the
dominant decay channel of the top quark, and in this context were
confirmed in \cite{cza90}.  The one-loop QCD corrections
decrease the rate of this decay by about $-8.4\%$ (for $m_t\approx
175$ GeV).

At the ${\cal O}(\alpha_s^2)$ level only the so-called BLM \cite{BLM}
corrections have been known \cite{smith,ac95a}.  Their numerical value
will be illustrated below.  Summation of these effects to all orders
has been discussed in \cite{Beneke:1995qe,Mehen:1997mw}.

The electroweak corrections to $t\to bW$ were evaluated in
\cite{den1} and were found to increase the decay rate by about
1.7\%.  This effect is almost canceled by accounting for the finite
width of the $W$, which decreases the rate by about $-1.5\%$
\cite{Jezabek:1993wk}.  In the present paper these both effects will
be neglected.

Present uncertainty in the theoretical prediction of the top quark
width is to a large extent connected with the unknown two-loop QCD
corrections \cite{Kuhn:1996ug} and the need for their complete
evaluation has been repeatedly emphasized.  This paper is devoted to
this calculation.  An exact computation of this effect would be very
difficult, but since the size of those corrections is expected to be
small, it is justified to make some approximations.  In the present
calculation we neglect the mass of the $W$; the error is expected to
be of the order $m_W^2/m_t^2 \approx 0.2$.\footnote{At the ${\cal
O}(\alpha_s)$ level, the difference between the QCD correction
calculated for $m_W=m_b=0$ and the correction calculated for physical
values of $m_W$ and $m_b$ is approximately $10\%$.}  Another, smaller
source of error is inherent in the method we employ in this
calculation.  As described below, we use the difference of the $t$ and
$b$ masses as an expansion parameter and calculate many (about 20)
terms of the series in this parameter.  An error of up to $6\%$ is
caused by the truncation of this series.  In any case, the accuracy of
our final result is completely sufficient for all applications which
can presently be contemplated.

In this paper we consider $W$ as stable.  Part of the effects of its
instability can be accounted for using the known results for the
${\cal O}(\alpha_s^2)$ corrections to its hadronic width.  There are
also ``non--factorizable'' corrections due to an exchange of gluons
between the top, or the $b$ quark resulting from its decay, and the
quarks produced in the $W$ decay. We expect these effects to be
additionally suppressed by  $\Gamma_W/m_W$.

There is also another application of the calculations reported in this
paper, related to the $b$ quark physics.  The inclusive decay width $b
\to u l \nu_l$ is of interest for the extraction of the CKM matrix
parameter $|V_{ub}|$.  The ${\cal O}(\alpha_s^2)$ corrections to $t
\to bW$ presented below, after simple modifications can be used
to obtain the corrections to the differential inclusive decay
width $\Gamma(b \to ul\nu_l)$ at the point where the invariant mass of
leptons vanishes. Thus, the results reported here can be used in the
future to estimate the second order QCD corrections to $b \to u
l\nu_l$.

Most of the results presented here have been derived using methods
described in \cite{maxtech} in the context of $b\to c$ decays.  In
addition, since the $t$ quark is much heavier than $b$, we have to
include three diagrams which result from ``non-planar'' interference
of $t\to Wbb\bar b$ amplitudes.  Their calculation is described in
some detail in the following section \ref{sec:ccc}.  In section
\ref{sec:twoLoop} we present the numerical results for various values
of the quark masses in the final state.  Our conclusions are given in
section \ref{sec:conclusions}.

\section{Non-planar diagrams $t\to bb\bar bW$}
\label{sec:ccc}
\begin{figure}[htb]
\hspace*{-6mm}
\begin{minipage}{16.cm}
\vspace*{13mm}
\[
\mbox{ 
\begin{tabular}{ccc}
\psfig{figure=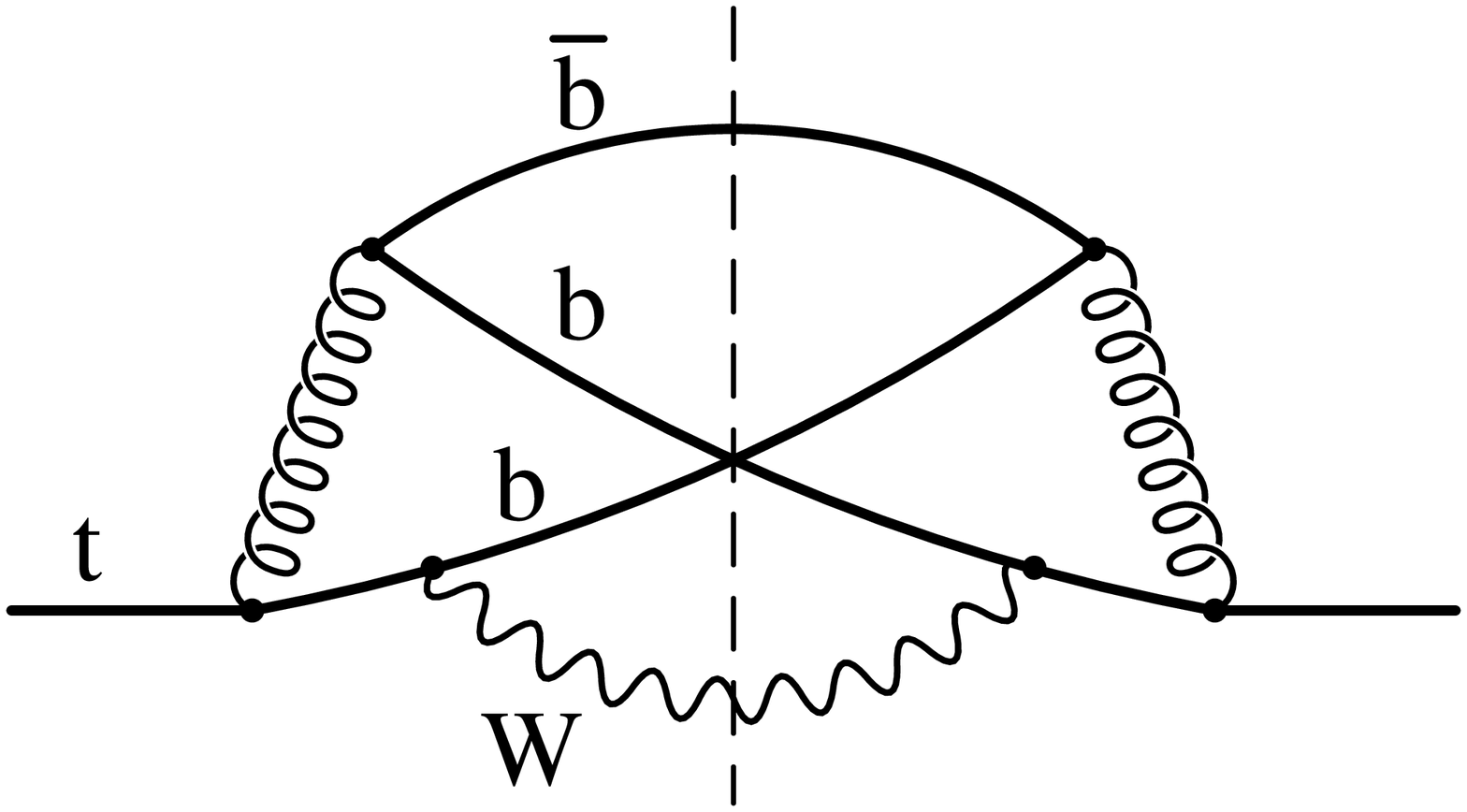,width=40mm,bbllx=72pt,bblly=291pt,%
bburx=544pt,bbury=530pt} 
& \hspace*{10mm}
\psfig{figure=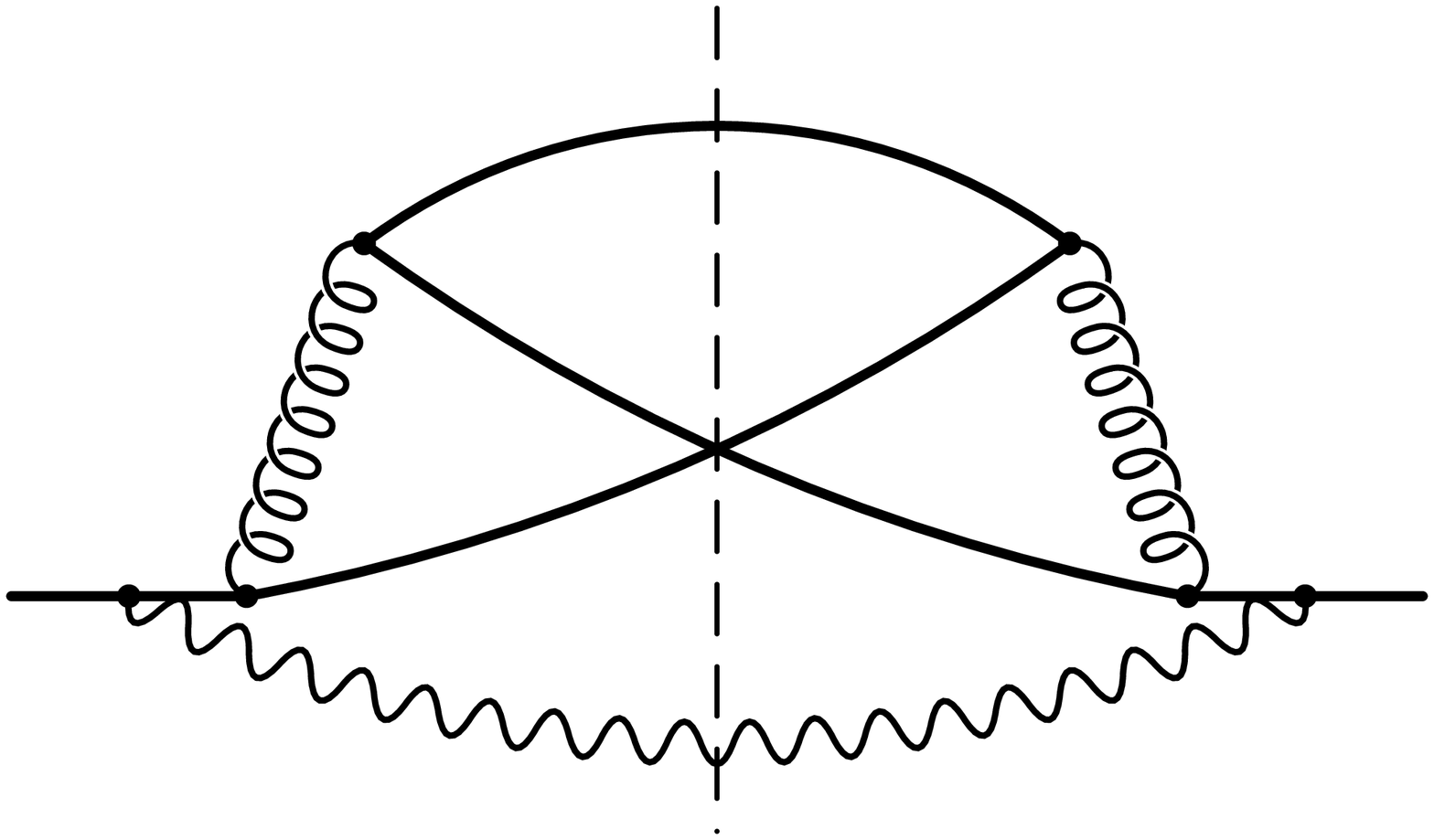,width=40mm,bbllx=72pt,bblly=291pt,%
bburx=544pt,bbury=530pt} 
& \hspace*{10mm}
\psfig{figure=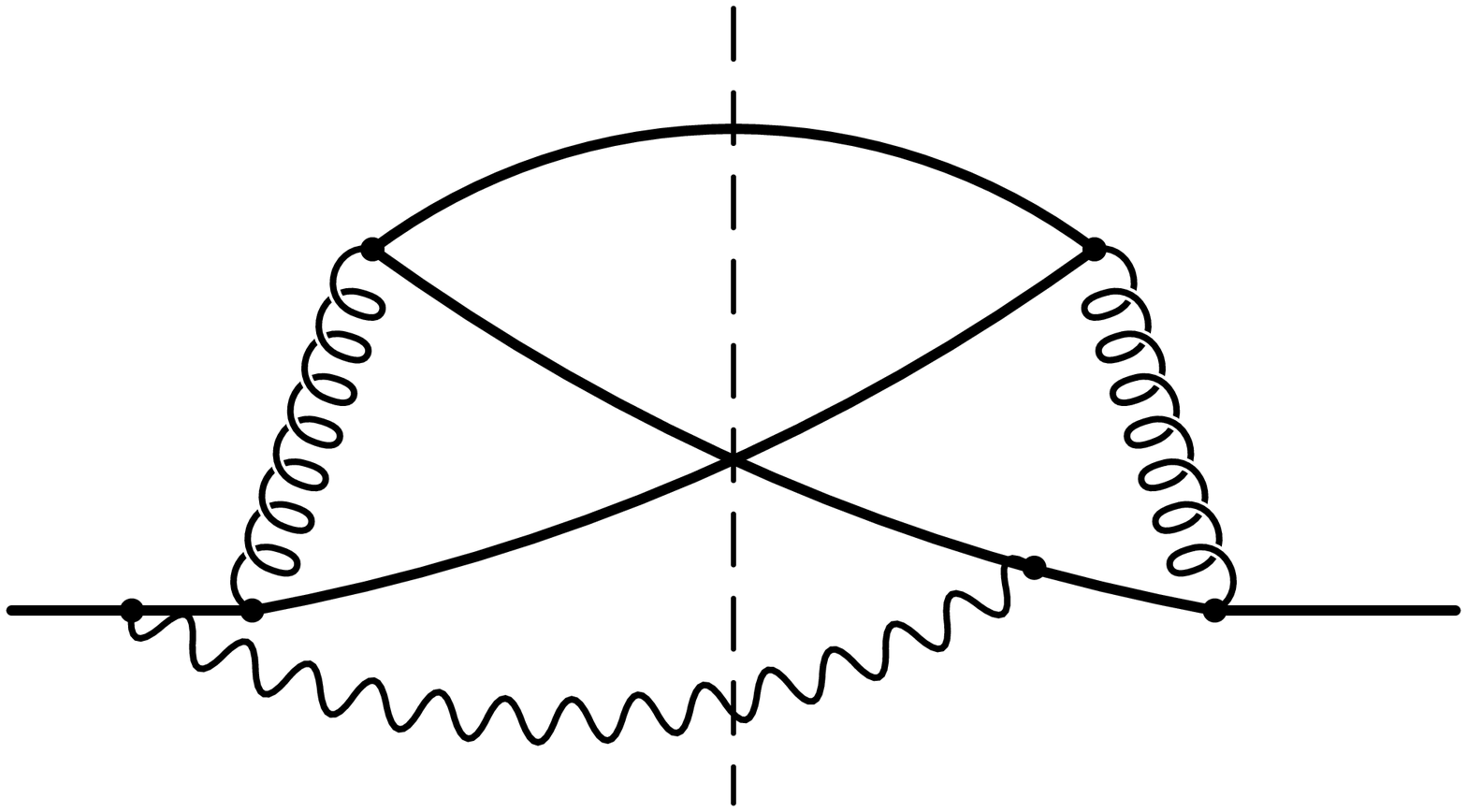,width=40mm,bbllx=72pt,bblly=291pt,%
bburx=544pt,bbury=530pt} 
\end{tabular}
}
\] \vspace*{0mm}
\end{minipage}
\caption{Non-planar diagrams contributing to the decay channel $t\to
Wbb\bar b$.}
\label{fig:c3}
\end{figure}

Since the top quark is significantly heavier than $m_W+3m_b$, the
decay $t\to bb\bar bW$ contributes to the width of the top at ${\cal
O}(\alpha_s^2)$.  The treatment of this decay channel is complicated
by the presence of non-planar interference diagrams shown in
fig.~\ref{fig:c3}.

In order to compute them, we expand around the
point $m_b=m_t/3$. The expansion variable is $\tilde\delta$ such that
\be 
m_b = \frac {m_t}{3}\;(1-\tilde\delta).  
\ee 

We introduce the following notations: $p_{1,2}$ the momenta of the
virtual gluons which later split into $b \bar b$ pairs, namely: $p_1 =
p_4+p_5$ and $p_2 = p_3+p_4$.  $p_3,p_5$ are the $b$ quark momenta and
$p_4$ is the antiquark momentum in the final state. $W$ denotes the
$W$-boson momentum, $t$ is the top quark momentum.  $p_7$ and $p_6$
are the momenta of the virtual top propagators: $p_6 = t - p_1$ and
$p_7 = t - p_2$. $b$ is the momentum of the virtual $b$--quark, $b =
t-W$.

All propagators can be expanded around the static limit,
because the  configuration around which we expand is the top
quark decaying into three $b$--quarks at rest. For instance,
the leading term in the expansion of 
the gluon propagator $1/p_1^2$ is $1/(4m_b^2)$. The virtual top
propagators give:
\be
\frac {1}{(t-p_1)^2 - m_t^2} \to -\frac {1}{4m_b(m_t-m_b)},
\ee
etc.  Therefore, no dependence on phase space variables 
remains in the expanded denominators and 
the integration over the phase space can be performed 
by successive factorization into two-particle phase spaces.

First we introduce the vector $H$ which combines $p_4$ and
$p_5$: $H=p_4+p_5$. We decompose $p_{4,5}$ into components parallel
and perpendicular (in four-dimensional sense)
to the $H$ direction. After  averaging over directions of
the perpendicular components this 
phase space integration gives the volume factor 
\be
R_2(H;p_4,p_5) = {\pi\over 2}\sqrt{1-{4m_b^2\over H^2}}.
\ee
Next, we combine $H$ and $W$ into $Q=H+W$. Since $W$ is masseless,
this phase space gives no square root function
\be
R_2(Q;H,W) = {\pi\over 2}{Q^2-H^2\over Q^2}.
\ee
We are then left with $t
= Q+p_3$ phase space and two integrations over $H^2$ and $Q^2$.  The
integration limits are $4m^2 < Q^2 < (M-m)^2$ and $4m^2 < H^2 <
Q^2$.  Also, the $Q,p_3$ phase-space is given by  K\"all\'en function,
the momentum of the $Q$ in the rest frame of $t$:
\be
R_2(t;Q,p_3) = {\pi\over 2m_t^2}
\sqrt{[(m_t-m_b)^2-Q^2][(m_t+m_b)^2-Q^2]}.
\ee
Finally, for $H^2$ and $Q^2$ we introduce variables $x_{1,2}\in (0,1)$,
defined by
$H^2 = 4m^2(1+\omega x_1x_2)$ and $Q^2 = 4m^2(1+\omega x_1)$.
From the limits on $H^2$ and $Q^2$  
we find
\be
\omega = {(m_t-m_b)^2\over 4m_b^2} - 1 \sim \tilde\delta
\ee
and the volume of the phase space $R_2(t;Q,p_3)$ is given by
\be
R_2(t;Q,p_3) = {4m_b\over \sqrt{3}m_t}\sqrt{\omega}\sqrt{1-x_1}
\sqrt{1-{{\tilde \delta}^2\over 4}-{x_1\tilde \delta\over 4}(4-\tilde \delta)},
\ee
where the last square root can be expanded in $\tilde\delta$.
Also the $\sqrt {1-4m^2/H^2}$ expressed in terms of $x_{1,2}$ becomes
proportional to $\sqrt {x_1 x_2}$.  As a result we obtain an
expression containing only half-integer or integer powers of $x_{1,2}$
which can be easily integrated.

\section{Two-loop corrections to top decay rate}
\label{sec:twoLoop}
In the limit of vanishing $W$ mass, 
the width of the decay $t\to bW$ is given by the formula (we neglect
terms ${\cal O}(\alpha_s^3)$)
\begin{equation}
\Gamma(t\to bW) =
\Gamma_0 \left[a_0+
\frac {\alpha_s(m_t)}{\pi}C_F a_1+ \left(\frac {\alpha_s}{\pi}\right)^2
C_F a_2 \right],
\label{eq:param}
\end{equation}
where $\Gamma_0 = \frac {G_F m_t^3}{8\sqrt{2}\pi}|V_{tb}|^2$.
In this paper we use $\alpha_s(m_t)$ determined in the $\overline{\rm
MS}$ scheme and $m_{t,b}$ are the pole masses.
$a_0 = (1-m_b^2/m_t^2)^3$ and $a_1$ is also known
in a closed analytical form \cite{jk2,sa92b}.  
In the limit of the
vanishing $b$ quark mass we have
\be
a_0 = 1, \qquad a_1 = {5\over 4} - {\pi^2\over 3}.
\ee
$a_2$ is the main new result of the present paper.  It can be divided
up into 4 gauge invariant parts 
\be
a_2 = \left( C_F-{C_A\over 2}\right) a_F
+ C_A a_A + T_RN_L a_L + T_R a_H,
\label{eq:a2}
\ee 
where $C_F= 4/3$, $C_A=3$, $T_R=1/2$, and $N_L$ is the number of quark
flavors whose masses can be neglected.  For the top quark decay we
take $N_L=5$.  $a_{F,A,L,H}$ are functions of the mass ratio of the
$b$ and $t$ quarks.  We have calculated them using an expansion around
the equal mass case.  The expansion parameter, $\delta\equiv
1-m_b/m_t$ is close to one, so that many terms of the series must be
calculated in order to obtain good accuracy.  Such expansion has
already been considered in our previous work
\cite{Czarnecki:1997hc,maxtech}. In those papers we have defined the
quantities $\Delta_{F,A,L,H}$.  Two coefficient functions in
eq.~(\ref{eq:a2}), $a_{A,L}$, coincide with
$\delta^3\Delta_{A,L}$.\footnote{There is a mistake in the formula for
$\Delta_A$ in \protect\cite{Czarnecki:1997hc,maxtech}.  In order to
obtain the correct result one has to add to $\delta^3\Delta_A$ given
there the following expression: $-{1\over
4}(\zeta_3-\pi^2/6)\sum_{n=1}^\infty
\beta^{n+4}n/[(n+1)(n+2)]$, with $\beta=\delta(2-\delta)$, expanded
in $\delta$ to desired order.  For $b\to cl\nu_l$ the resulting change
is insignificant; the magnitude of $\Delta_A$ is decreased by less
than 3\%.}  For the purpose of the present application we have
more than doubled  the number of terms in the  expansion.  $a_H$ is
analogous to $\delta^3\Delta_H$, except that it now contains
contributions of the top quark loop only.  The $b$ quark loop and part
of real $b\bar b$ pair radiation is accounted for by adjusting $N_L$.
There is, however, part of real $b\bar b$ pair production which cannot
be described in this way.  It corresponds to the non-planar
interference diagrams shown in fig.~\ref{fig:c3} and is included in
$a_F$, which without these diagrams would have coincided with
$\delta^3\Delta_F$.  These diagrams, contributing to the decay $t\to
Wbb\bar b$, have no analogs in the $b-c$ transitions considered in
\cite{Czarnecki:1997hc,maxtech}.

\begin{figure}[htb] 
\hspace*{-6mm}
\begin{minipage}{16.cm}
\vspace*{13mm}
\[
\mbox{ 
\begin{tabular}{cc}
\psfig{figure=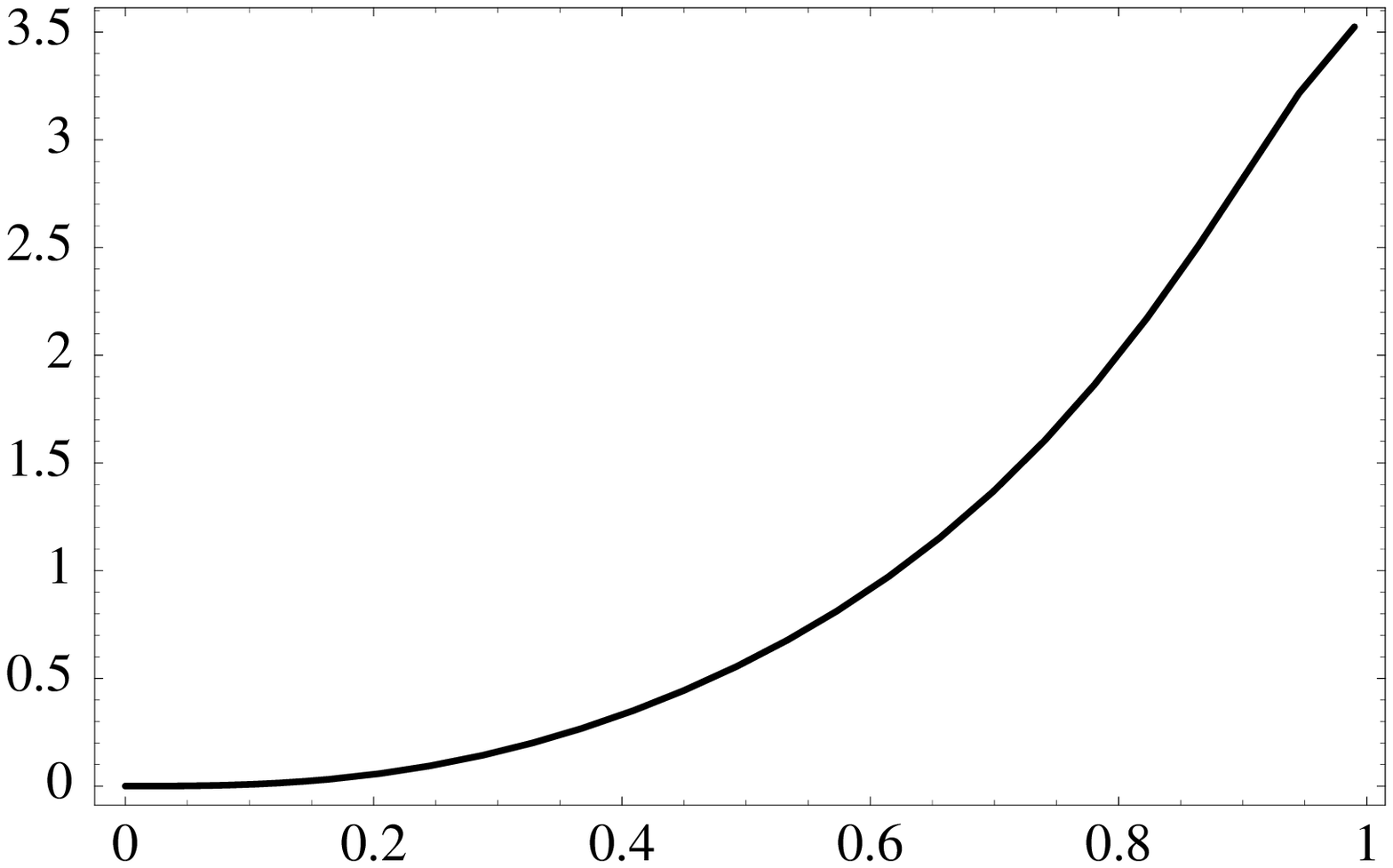,width=70mm,bbllx=72pt,bblly=291pt,%
bburx=544pt,bbury=530pt} 
& \hspace*{6mm}
\psfig{figure=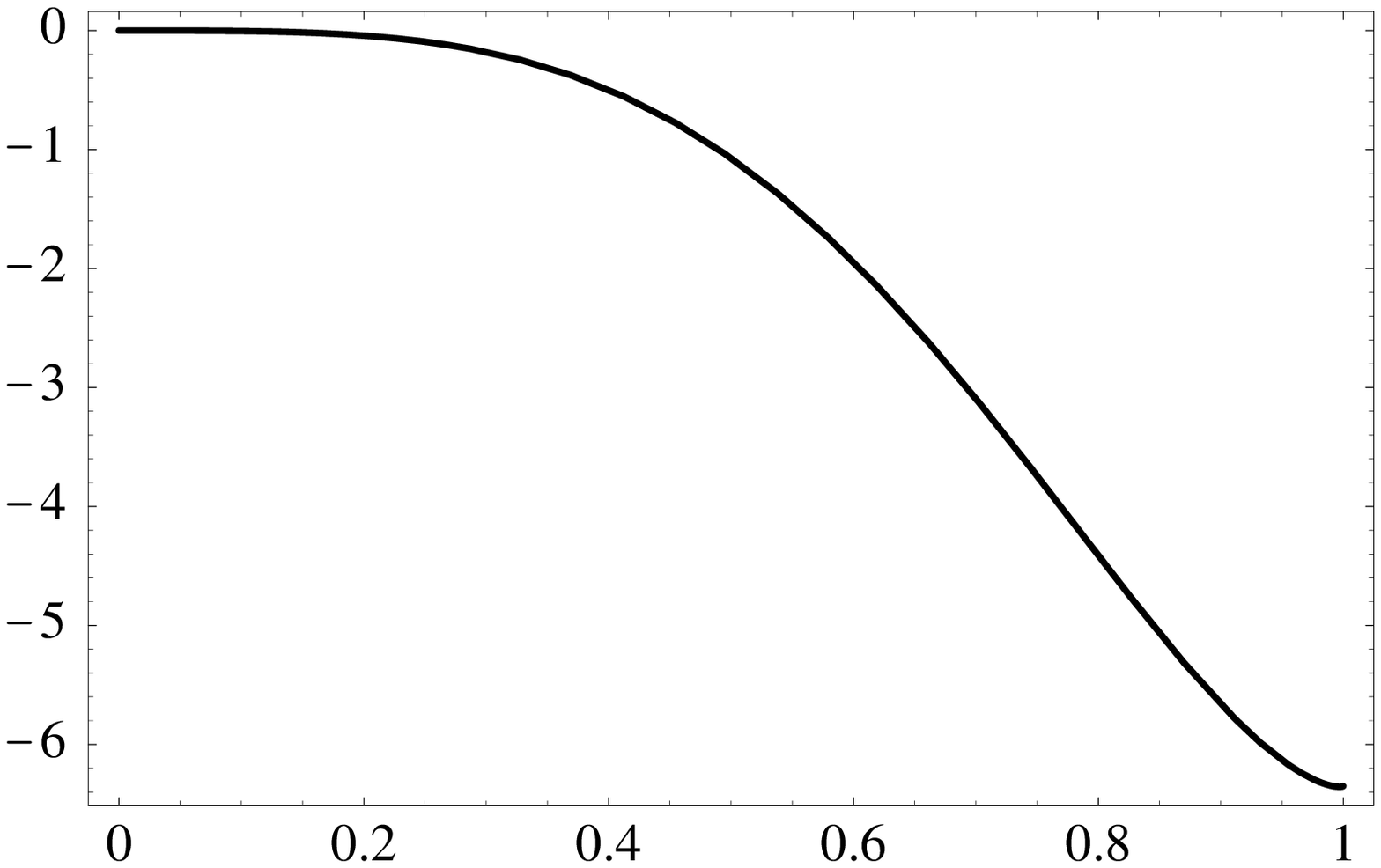,width=70mm,bbllx=72pt,bblly=291pt,%
bburx=544pt,bbury=530pt} 
\\[5mm]
(a) & \hspace*{6mm} (b)
\\[5mm]
\psfig{figure=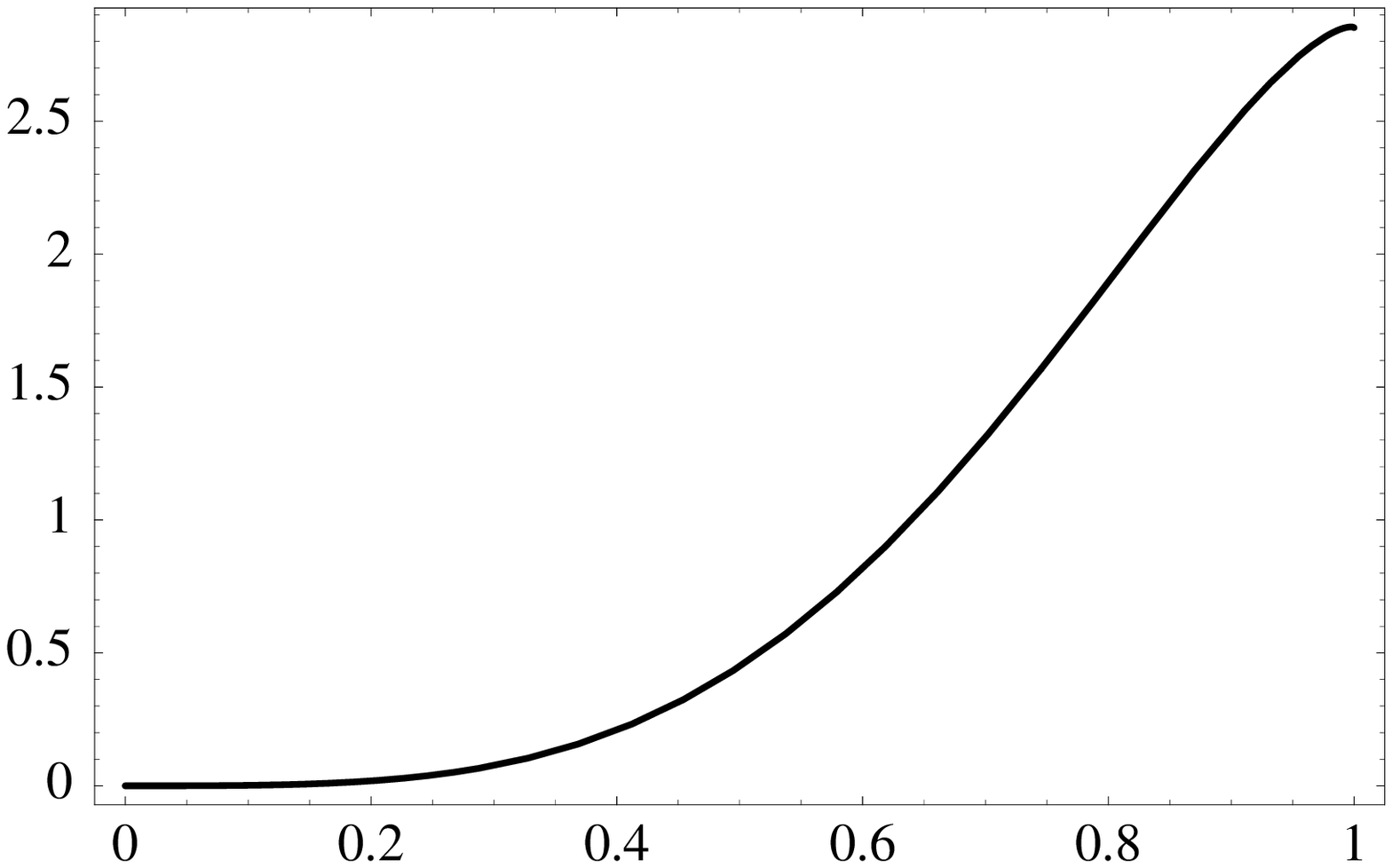,width=70mm,bbllx=72pt,bblly=291pt,%
bburx=544pt,bbury=530pt} 
& \hspace*{6mm}
\psfig{figure=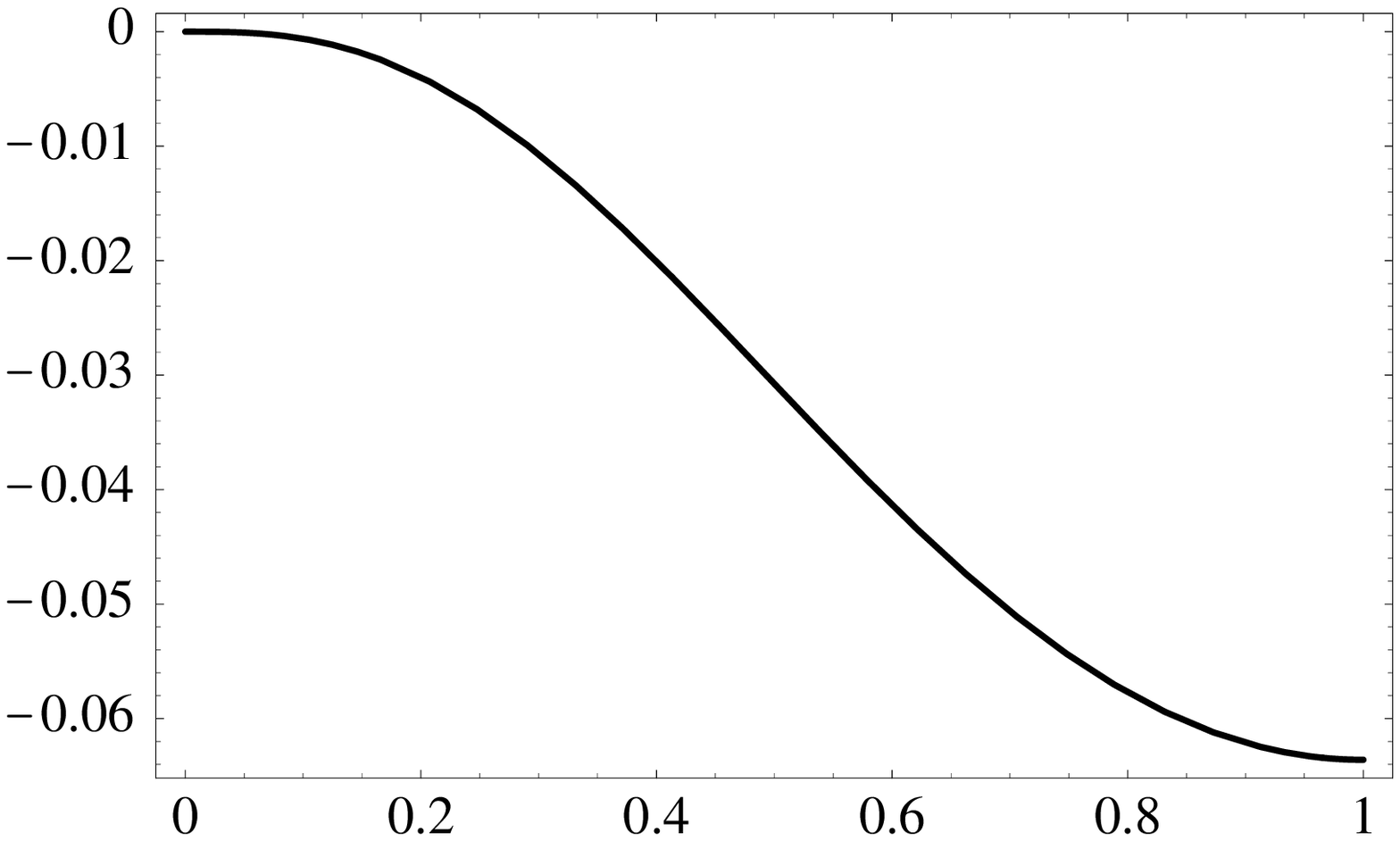,width=70mm,bbllx=72pt,bblly=291pt,%
bburx=544pt,bbury=530pt} 
\\[5mm]
(c) & \hspace*{6mm} (d)
\end{tabular}
}
\] \vspace*{0mm}
\end{minipage}
\caption{Plots of the coefficient functions defined in
eq.~(\protect\ref{eq:a2}): (a) $a_F$, (b) $a_A$, (c) $a_L$, (d)
$a_H$.  The variable on the horizontal axis is $m_b/m_t$.}
\label{fig:plots}
\end{figure}

For $m_b\to 0$,  $a_L$ was obtained numerically in \cite{smith}, and  is now
known analytically  in this limit \cite{ac95a}:
\be
a_L(m_b=0) = -{4\over 9}+{23\over 108}\pi^2 +\zeta_3, \qquad \zeta_3
\simeq 1.2020569. 
\label{eq:aL}
\ee
The dependence of the coefficient functions $a_i$ on $m_b/m_t$ is
shown in fig.~\ref{fig:plots}.
For  $m_b/m_t \approx 4.8/175$ we find
the following numerical values:
\be
\mbox{For } m_b/m_t &=& 4.8/175:
\nonumber \\
 a_F &=& 3.4(2), \qquad 
a_A = -6.28(2), \qquad
\nonumber \\
a_L&=&2.83(4),\qquad 
a_H = -0.06349(1).
\ee
We note that $a_F$ has the largest error bar.  This is because there
are two separate series contributing to it: in addition to the series
in $\delta=1-m_b/m_t$ there is also the non-planar contribution
expanded in $\tilde\delta=1-3m_b/m_t$.  Both these series are
separately divergent in the limit $m_b\to 0$ like $\ln^3(m_b)$.  For
this reason their reliable estimate near this limit is difficult and
we have assigned a conservative error bar to the result.  
On the other hand, because of the color factors in eq.(\ref{eq:a2}),
the final result for $a_2$ is not very  sensitive 
to the relatively large error in $a_F$.

The expressions we have obtained for $a_{F,A,L}$ contain the first and
second powers of $\ln(2\delta)$.  It turns out that the convergence of
these series is improved if we rewrite these terms as
$\ln(2\delta)=\ln\beta - \ln(1-\delta/2)$, with
$\beta=\delta(2-\delta)$ and expand $ \ln(1-\delta/2)$ in $\delta$.
Since $\beta$ approaches 1 much faster than $\delta$, the remaining
logs are smaller.  Because of the many terms which had to be evaluated
for the coefficient functions $a_i$, the results are rather lengthy
and not suitable for publication in a journal.  However, they can be
obtained from the authors upon request.

In the future, it is likely that the corrections to $t\to bW$ will be
calculated analytically in the point $m_b=0$.  We therefore list here
our estimates for the numerical values of $a_i$ in this point:
\be
\mbox{For } m_b/m_t &=& 0:
\nonumber \\
a_F &=& 3.5(2), \qquad
a_A = -6.35(7),  \qquad
\nonumber \\
a_L&=& 2.85(7), \qquad 
a_H = -0.06360(1).
\ee
The value of $a_L$ found from our expansion is in agreement with the
exact value obtained from eq.~(\ref{eq:aL}),
$a_L(m_b=0)=2.859473\ldots$.

Let us now discuss the numerical value of the correction
in the case of the massless $b$ quark.
For the purpose of discussion it is convenient to separate
the BLM \cite{BLM} and the non-BLM corrections to the decay rate.
The BLM corrections follow immediately from the results
of \cite{smith,ac95a} and in present notations are given by:
\be
a_2^{\rm BLM} = -T_R\left(\frac {33}{2} - N_L \right)a_L.
\ee
Using $N_L=5$, we obtain:
\be
a_2^{\rm BLM}(m_b=0) = -16.442\ldots.
\ee
This is to be compared with the complete result for $a_2$,
\be
a_2(m_b=0) = -12.5(4).
\ee
The complete ${\cal O}(\alpha_s^2)$ corrections are smaller by 24\%
than the BLM estimate.  This difference is somewhat larger than in
$b\to c$ decays \cite{Czarnecki:1997hc,maxtech}.

In any case, numerically the second order corrections appear
to be very moderate:
\be
\Gamma_t(m_b/m_t=0) = \Gamma_0 \left( 1-0.866\; \alpha_s(m_t) 
 - 1.69(5)\; \alpha_s^2  \right).
\label{gam0}
\ee

%The BLM-based estimate of the ${\cal O}(\alpha_s^2)$ correction
%would deliver a number close to $2.5$ per cent on the other hand.

For $m_b/m_t=4.8/175$ we find
\be
a_2(m_b/m_t=4.8/175) &=& -12.4(3),\qquad 
a_2^{\rm BLM}(m_b/m_t=4.8/175) = -16.3(2);
\nonumber\\
\Gamma_t(m_b/m_t=4.8/175) &=&  0.998 \times \Gamma_0 \left(1
- 0.857\alpha_s(m_t)-1.68(4)\alpha_s^2\right).
\label{gam1}
\ee
Using $\alpha_s(m_t) = 0.11$, we find that the ${\cal O}(\alpha_s^2)$
corrections to the top decay width are very close to 2\%.

Finally, we list here also the values obtained for $\delta=0.7$, or
$m_b/m_t=0.3$, which after replacement $t,b\to b,c$ are 
relevant for the differential width of $b\to
c+\mbox{leptons}$, when the leptons are emitted parallel to each
other, i.e. with a zero invariant mass:
\newpage
\be
\mbox{For } m_b/m_t &=& 0.3:
\nonumber \\ 
a_F &=& 1.37261(2), \qquad
a_A = -3.09498(4),  \qquad
\nonumber \\
a_L&=& 1.31091(5),  \qquad 
a_H = -0.05061395(1).
\ee
At $m_b/m_t=0.3$ one has to account for the finite $b$ mass in the
virtual loop. This amounts to replacing $a_H$ by $\tilde{a}_H$, which
corresponds to $\delta^3\Delta_H$ in \cite{Czarnecki:1997hc,maxtech}.
At $\delta=1-m_b/m_t=0.7$ we find
\be
\tilde{a}_H = a_H + (b\mbox{-loop}) \simeq -0.0506+0.6333 = 0.5827.
\ee

\section{Conclusion}
\label{sec:conclusions}
We have presented a calculation of the ${\cal O}(\alpha_s^2)$
corrections to the decay width of the top quark $\Gamma(t \to b W)$ in the
limit $m_t \gg m_W$. For $\alpha_s(m_t)=0.11$, we found (see
eqs. (\ref{gam0}, \ref{gam1})) that the second order QCD corrections
decrease the value of the top width by 2\%.

To perform this calculation, we used the methods described in detail
in \cite{maxtech}. Also, since the process $t \to b\bar b b W$
contributes to the top decay width at order ${\cal O}(\alpha _s^2)$,
we had to develop a new technique to deal with the ``non-planar''
interference diagrams.
 
The decay rate of the top quark is proportional to the third power of
its mass. In our calculations we used the pole mass of the top
quark. It has been demonstrated \cite{Beneke:1995qe} that the
convergence of the perturbation series is improved if one parametrizes
the width formula in terms of the $\overline {\rm MS}$ mass.  We have
not performed this reparametrization here, since the second order
corrections are small even if the pole mass is used.  We note,
however, that employing the $\overline {\rm MS}$ mass is likely to
decrease them even further.

With the electroweak corrections and effects of the $W$ width on
$\Gamma_t$ known, our result provides the last missing ingredient in
predicting the top decay width in the SM with accuracy $\sim 1\%$.
The remaining uncertainty is presently dominated by the top quark
mass.  It is difficult to say at the moment whether or not the
experimental determination of the top quark width can be performed
with a comparable precision.  A future muon collider might be
the best place for such measurements.

\section*{Acknowledgments}

We thank M. Je\.zabek and J. H. K\"uhn for reading the manuscript and
helpful comments. This work was supported in part by DOE under grant
number DE-AC02-98CH10886, by BMBF under grant number BMBF-057KA92P,
and by Gra\-duier\-ten\-kolleg ``Teil\-chen\-phy\-sik'' at the
University of Karlsruhe.

\newpage

%\bibliographystyle{../../pro/tex/prsty}
%\bibliography{../../pro/tex/phd}

\end{document}